\documentclass[%
 preprint,
 amsmath,amssymb,
 aps, physrev,
]{revtex4-2}

\usepackage{graphicx}% Include figure files
\usepackage{dcolumn}% Align table columns on decimal point
\usepackage{bm}% bold math
\usepackage{xcolor}

\begin{document}

\title{\textbf{ Strong altermagnetism and topological features in a two-dimensional van der Waals heterostructure via broken time reversal symmetry} 
}% 

\author{Anil Kumar Singh$^1$}
 \altaffiliation{$^1$Department of Physics, Tezpur University (Central University), Tezpur-784028, India.}%Lines break automatically or can be forced with \\
\author{Pritam Deb$^1$}%
 \email{Contact author: pdeb@tezu.ernet.in}
\affiliation{%
 Authors' affiliations\\
  $^1$Department of Physics, Tezpur University (Central University), Tezpur-784028, India.
}%

\begin{abstract}
The advent of altermagnetism, a new phase of magnetism, has garnered significant interest due to its extraordinary spin-polarized electronic bands despite zero net magnetization. Such spin-symmetry-guided robust non-relativistic alternating spin splitting in a compensated collinear magnet presents a novel platform for magnetotransport and nontrivial topology. Predominantly, altermagnetic behavior is observed in bulk magnetic materials upon incorporating external perturbations. However, van der Waals heterostructures can offer exceptional flexibility in tailoring their various emerging properties without the need for external perturbations in the two-dimensional regime. Here, an unconventional time reversal symmetry breaking with sizeable spin splitting via broken space-spin symmetry (\textit{P$\mathcal{T}$}) has been demonstrated in an antiferromagnet/nonmagnet vdW heterostructure. The lifted Kramer's degeneracy alongwith spin-orbit interactions result in non-zero Berry curvature, contributing to the outstanding magnetotransport with a large value of anomalous Hall conductivity ($\sim 732.9\ {\mathit{\Omega}}^{-1}{cm}^{-1}$). The presence of relativistic spin-orbit interactions in addition to predominant non-relativistic effects governed spin-momentum locking with a weak out-of-plane Dzyaloshinskii-Moriya interaction, which induces small spin canting. Further, the lowest magnetic anisotropy energy confirms collinear antiferromagnetic ground state. In particular, a nontrivial topology is observed along the surface [001], which is confirmed by the non-zero Chern number. This study provides a novel approach to realize strong altermagnetism in broken space-spin symmetry systems and fosters emergent transport behaviors.
\end{abstract}

\maketitle

\section{\textbf{INTRODUCTION}}
Compensated collinear magnets, a major breakthrough discovery of recent time, with unconventional time reversal symmetry (TRS) breaking are of great interest for both fundamental research and spintronics device applications. This spin-symmetry-guided magnetic phase due to diminishing stray fields and large alternating spin splitting in momentum space with zero net magnetization demonstrates unprecedented magneto-transport, anomalous Nernst effect, nontrivial band topology, spin current, etc. \cite{1,2,3,4,5,6,7,8} The magnetic order compensated via crystal symmetry leads to various emergent phenomena such as spin-charge interconversion, magneto-optical responses, and switching through its unique intrinsic properties. Additionally, lifting spin degeneracy in altermagnets via spin-lattice coupling exhibits an exotic electronic structure and provides great control of spin degree of freedom. 

This emerging third magnetic phase can be described by the nontrivial spin-space group as $\mathbf{R}_S^{III} = [E \parallel \mathbf{H}] + [C_2 \parallel \mathbf{G} - \mathbf{H}]$, where $\mathrm{E}$, $\boldsymbol{\mathrm{G}}$ and $\boldsymbol{\mathrm{H}}$ represents identity element, crystallographic Laue group and halving subgroup of $\boldsymbol{\mathrm{G}}$ respectively \cite{9}. The coset $\boldsymbol{\mathrm{G}}\boldsymbol{\mathrm{-}}\boldsymbol{\mathrm{H}}$, which includes real space rotational transformation, connects the inequivalent magnetic sublattices with opposite magnetic moments. The requirement for rotational or mirror symmetries to break space-time inversion symmetry (\textit{P$\mathcal{T}$}) imposes constraints on the space groups that exhibit altermagnetism (AM) \cite{9, 10, 11, 12}. Consequently, only specific magnetic space groups host altermagnetic spin splitting. Moreover, the discovery of two-dimensional (2D) magnetism unveils fascinating properties, motivating the pursuit of realizing this third magnetic phase, a new class distinct from ferromagnets and antiferromagnets \cite{13, 14}. The low-dimensional altermagnetic behaviors can stimulate surface dependent distinct electronic characteristics. 

Until now, the realization of AM has been limited to bulk (3D) magnetic materials, which hinders its full potential for applications \cite{1}. For example, researchers have established AM in theoretically designed and experimentally synthesized bulk RuO${}_{2}$ \cite{3,5,15,16,17,18,19}. An antiferromagnetic spin splitting is also observed in bulk magnetic materials under external perturbations \cite{20,21}. Further, AM has been predicted in pristine atomic crystals and homo-bilayers upon the application of external perturbations such as electric field, janusization, crystal distortion, twisting, and doping \cite{11,22,23,24,25,26}. Subsequently, a general stacking theory (GST) has been proposed for the functionalization of AM in twisted bilayer systems (e.g. CrN, 2H-NiCl${}_{2}$, 1T-NiCl${}_{2}$ and CrI${}_{3}$) \cite{22}. Moreover, a microscopic model has been presented for AM in 2D antiferromagnets and its tunability via magnon-electron coupling \cite{23}. However, the reliance on external disruptions to realize AM in low-dimensional materials hinders the fabrication of 2D systems due to instability and complexity, while also altering their intrinsic properties. Further, the presence of magnetic interactions such as the Heisenberg exchange and Dzyaloshinskii-Moriya interaction (DMI) plays a pivotal role in stabilizing ground state magnetic configuration. The large value of isotropic Heisenberg exchange can lead to collinear ferromagnet or antiferromagnet, whereas strong antisymmetric DMI can induce large spin canting in the presence of SOC \cite{27, 28}. In this context, engineering a van der Waals (vdW) heterostructure consisting of an in-plane antiferromagnet (AFM) and nonmagnetic (NM) atomic layer with minimal lattice mismatch ($\mathrm{\sim}$1\%) could enable the realization of non-relativistic spin splitting through broken \textit{P$\mathcal{T}$} symmetry, while preserving the point group symmetry. VdW (AFM/NM) heterostructures also provide utmost control, stability, and proximity-induced spin-orbit coupling \cite{29,30,31}.

In the present study, strong AM has been demonstrated in vdW heterostructure having broken inversion and time reversal symmetry. The designed bilayer Mn${}_{2}$P${}_{2}$S${}_{6}$/Mo${}_{3}$Te${}_{6}$ vdW heterostructure exhibits large alternating spin splitting along the M1-MC-M2 high-symmetry path with zero net magnetization. Symmetry analysis reveals unconventional TRS breaking, which can be attributed to the intrinsically broken \textit{P$\mathcal{T}$} symmetry due to the presence of a nonmagnetic monolayer of 1H- Mo${}_{3}$Te${}_{6}$. This leads to the emergence of remarkable anomalous Hall conductivity (AHC) and surface-dependent band topology. On the contrary, the constituent semiconducting 1H-Mn${}_{2}$P${}_{2}$S${}_{6}$ monolayer host in-plane AFM configuration, in which the presence of inversion symmetry enforces spin degeneracy and preserves TRS \cite{11,24,32,33,34}. Accordingly, pristine AFM with zero magnetization and spin degeneracy could not possess such peculiar electronic and magnetic properties. Hence, the design of a vdW heterostructure offers a distinctive approach to the functionalization of AFM into AM to facilitate low-dimensional spin-correlated phenomena.

\section{\textbf{COMPUTATIONAL DETAILS}}
First principle-based density functional theory (DFT) calculations have been carried out to achieve ground state electronic, magnetic and structural properties of the designed Mn${}_{2}$P${}_{2}$S${}_{6}$/Mo${}_{3}$Te${}_{6}$ vdW heterostructure. The weak interlayer vdW force has been embedded via ‘DFT-D’ vdW correction method \cite{35}. Moreover, Hubbard correction (DFT+U) is incorporated to reflect onsite interactions by considering U=4 for Mn-d orbitals, which interprets the electronic structure more accurately \cite{36}. In this regard, QUANTUM Espresso Simulation package has been implemented to execute self-consistent calculations based on Kohn-Sham formalism \cite{37}. The self-consistent calculations have been performed under optimized environment, which includes a dense ‘26×26×1’ Monk-horst and Pack K-point mesh \cite{38}. These calculations have been accomplished with a kinetic energy value of 408 eV and the image interactions between the layers is avoided by considering a vacuum of 30 {\AA}. The electron-ion dynamics in the heterostructure is taken into account using projector augmented pseudopotentials \cite{39}. It is worthwhile to mention that the threshold energy and Hellmann-Feynman force criterion to achieve convergence is considered to be $1 \times 10^{-9}$ eV and $1 \times 10^{-4}$ Ry/Bohr respectively. The exchange interactions are considered via Perdew-Burke-Ernzerhof exchange functional with generalized gradient approximation \cite{40}. Following the electronic property calculations, the magneto-transport properties, the surface-dependent altermagnetic and topological properties have been calculated using Wannier90 code and WannierTools \cite{41,42,43,44}. The wannierized bandstructure with spin-orbit coupling (SOC) is calculated by projecting Mn-d, Mo-d, P-p, S-p and Te-p orbitals. The observed SOC motivates to calculate Berry curvature at -0.936 eV, since the anomalous Hall conductivity is found to be maximum at this energy value.Further, surface states, spin texture, Fermi arc and nontrivial topological properties such as Wannier charge center (WCC) are calculated by employing WannierTools.

\section{\textbf{RESULTS AND DISCUSSIONS}}
The engineering of a magnetic vdW heterostructure consisting of monolayers with minimal strain widens the scope of low-dimensional magnetism and spintronics applications \cite{45, 46, 47}. Here, the modeled vdW heterostructure consists of ML-MnPS${}_{3}$, an in-plane antiferromagnet, and ML-MoTe${}_{2}$, which serves as a nonmagnetic semiconductor substrate. The in-plane antiferromagnetic ML-MnPS${}_{3}$ with lattice constant $a_{MnPS_3}=6.12\ \textrm{\AA}$, belongs to the high-symmetry space group 162, P$\overline{3}$1m, which possesses manifold crystal symmetry. Noticeably, GST predicted that atomic layers with P$\overline{3}$1m layer group symmetry intrinsically host AM in their twisted homo-bilayer structure \cite{22}. Furthermore, the point group symmetry $D_{3h}$, illustrates the presence of horizontal mirror symmetry alongside three-fold rotational symmetry. The presence of symmetry operations (i.e., inversion center, mirrors and rotation) map electronic charges of opposite spin sublattices on each other that can preserve the $\textit{P} \mathcal{T}$ symmetry (spin degeneracy and TRS) leading to in-plane AFM [Fig. 1(a)]. On the other hand, ML-MoTe${}_{2}$ is a nonmagnetic vdW material with semiconducting band gap in its hexagonal phase. It consists of three atomic layers in which the Mo atom is sandwiched between two Te atoms, with lattice parameters $a_{MoTe_2}=3.56\ \textrm{\AA}$ and $\alpha = \beta = 90^\circ, \ \gamma = 120^\circ$. Noticeably, the lattice parameter $\sqrt{3}a_{MoTe_2} =6.17\ \textrm\AA$  of MoTe${}_{2}$ matches that of ML-MnPS${}_{3}$ with a negligible lattice mismatch of 0.8\%. This motivates the design of a Mn${}_{2}$P${}_{2}$S${}_{6}$/Mo${}_{3}$Te${}_{6}$ vdW heterostructure as shown in Fig. 1(b, c), which allows the constituents to endure their intrinsic electronic and magnetic properties while enabling the emergence of new magnetic phase. The designed heterostructure shows superior thermal stability with minimal energy [blue color] and temperature [red color] fluctuations, as predicted by \textit{ab initio} molecular dynamics simulation [refer to Fig. 2]. In fact, the substrate ML-MnPS${}_{3}$ is intended to provide stability and break the $\textit{P} \mathcal{T}$ symmetry of the system. In addition, it breaks the horizontal mirror symmetry of the AFM layer, which is essential for AM. This space-inversion symmetry breaking transforms the vdW system into a low-symmetry crystal structure. Figure 1(b) displays the atomic arrangements of the crystal structure with optimized interlayer separation, which is calculated to be 3.3 {\AA}. It is observed that the magnetic Mn atoms and nonmagnetic P atoms reside on top of the center of hexagonal MoTe${}_{2}$, whereas Se atoms are positioned on top of Mo and Te atoms as shown in Fig. 1(c).

\begin{figure*}[th!]     
\centering           
\includegraphics[width=16cm,height=8cm]{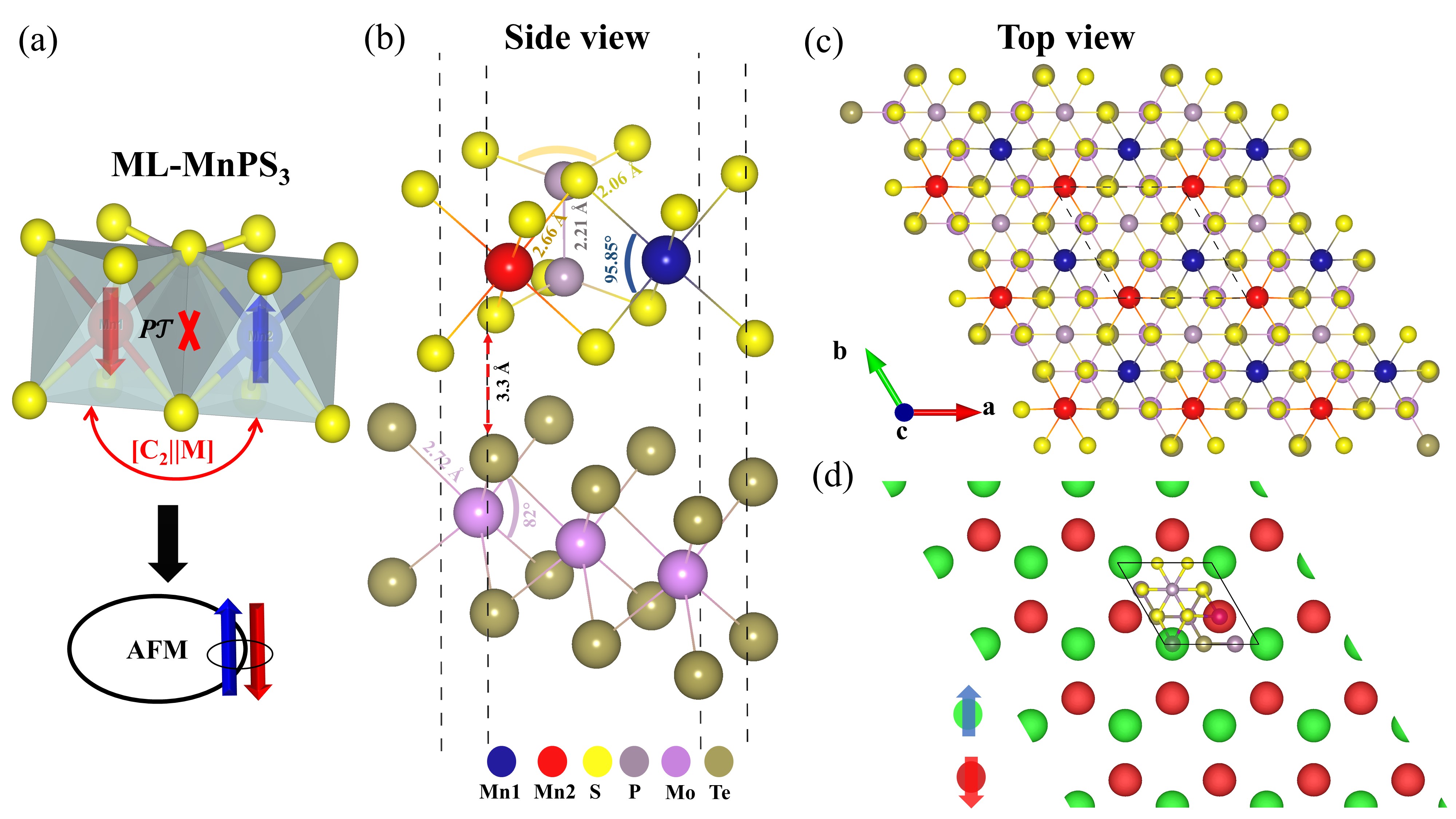}
\caption{\label{Fig:wide}  Crystal structure and atomic configurations. (a) Schematic representation of crystal symmetry of in-plane antiferromagnetic ML-MnPS${}_{3}$, (b, c) atomic configurations of Mn${}_{2}$P${}_{2}$S${}_{6}$/Mo${}_{3}$Te${}_{6}$ vdW heterostructure with side view and top view respectively, (d) spin densities of Mn atoms [green and red balls represent magnetic moment along $+Z$ and $-Z$ respectively]. }   
\end{figure*}

Importantly, the unit cell consists of opposite sublattices of two Mn atoms with opposite magnetic moment orientations, leading to zero net magnetization. Figure 1(d) represents the spin densities of these two magnetic atoms with AFM configuration. However, the \textit{P$\mathcal{T}$} symmetry breaking compels the broken TRS of the AFM layer despite the absence of net magnetization. This leads to the emergence of a new class of compensated collinear magnetic phases termed altermagnet (a third phase of magnetism in addition to FM and AFM). Hence, the electronic structure of the system is also modified to reflect the observed breaking of TRS.

Typically, the bandstructure of compensated collinear antiferromagnet upholds TRS by virtue of space-inversion symmetry, which connects the opposite spin sublattices. This governs the overlapping of electronic states with opposite spins and imposes spin-degeneracy in momentum space. However, the lifting of spin-degeneracy in compensated magnets, driven by decoupled space-spin symmetry, gives rise to an unconventional magnetic phase with strong alternating spin splitting. The appearance of alternating spin splitting along the high-symmetry K-path flourishes groundbreaking spin-correlated attributes. Hereof, we calculated the electronic structure of the heterojunction as shown in Fig. 3. 

\begin{figure*}[th!]     
\centering           
\includegraphics[width=12cm,height=6cm]{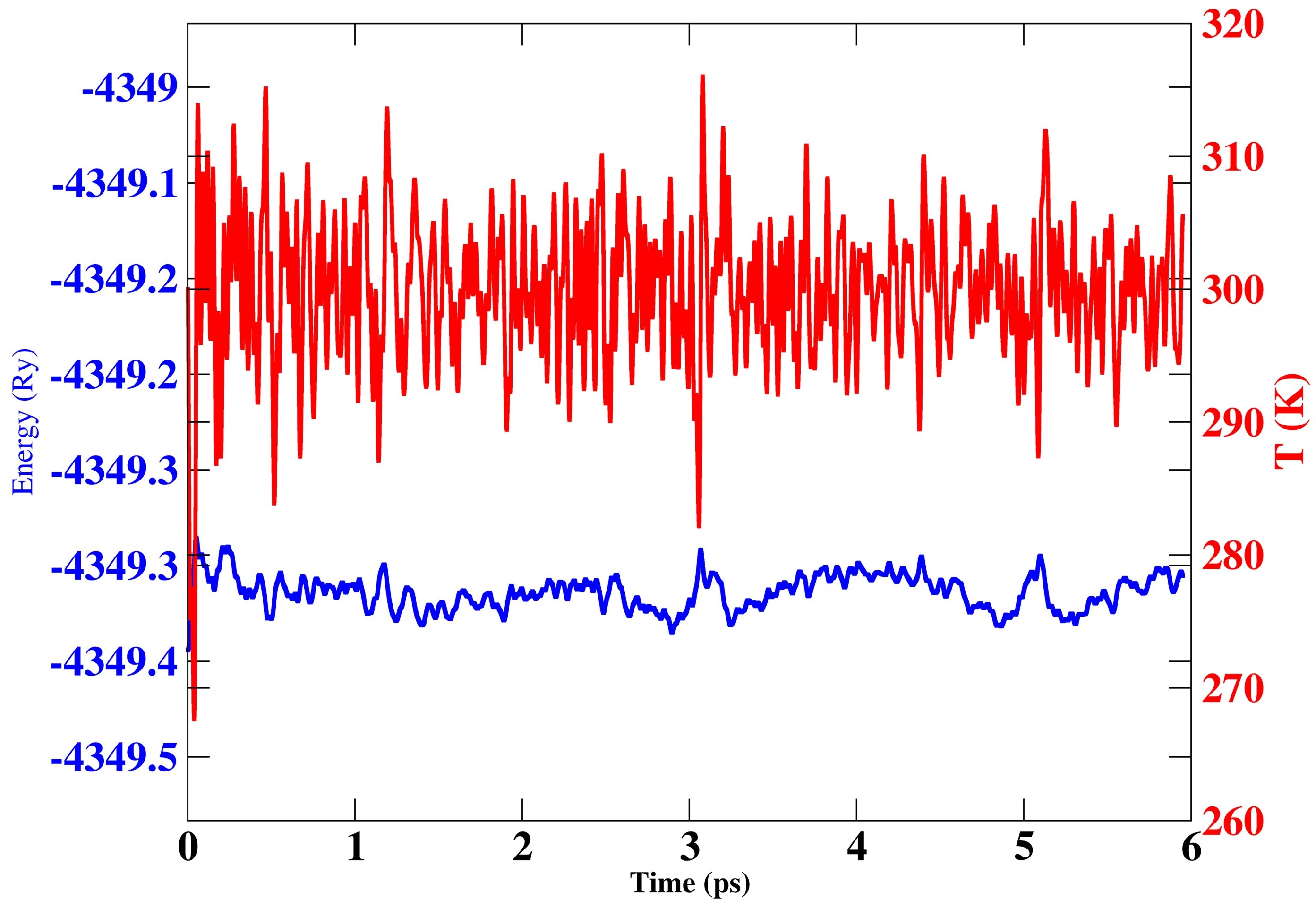}
\caption{\label{Fig:wide} Thermal stability of the designed Mn${}_{2}$P${}_{2}$S${}_{6}$/Mo${}_{3}$Te${}_{6}$ altermagnet via \textit{ab initio} molecular dynamics simulation. }
\end{figure*}

Figure 3(a, b) shows that the bandstructure exhibits lifted spin-degeneracy with alternating spin splitting along M1-MC-M2 high-symmetry path in both valence band (VB) and conduction band (CB) region. It is observed that the MC-point preserves spin degeneracy, which may be due to the fact that MC-point is invariant under real-space transformations \cite{1}. However, the electronic states around MC-point, i.e., along MC$\mathrm{\to}$M1/M2, exhibit spin splitting, where the spin-up states along MC$\mathrm{\to}$M1 match with spin-down states along MC$\mathrm{\to}$M2, and vice-versa. Additionally, the bandstructure as in Fig. S1 (Sec. I), reflects the indirect band gap nature with valence band maximum (VBM) and conduction band minimum (CBM) located at M1/M2 and MC high-symmetry points respectively \cite{48}. The bandstructure features both flat bands and dispersive parabolic bands, leading to distinct local and global transport properties of the charge carriers. Specifically, the electronic states are found to be flat near the Fermi level but become dispersive at higher energy region. Interestingly, the alternating spin splitting is found to be linear in the vicinity of MC-point, which may give rise to nontrivial topological characteristics. This compensated collinear magnetic configuration is further confirmed by the large Heisenberg exchange value 15.8 meV with a positive sign ($\boldsymbol{J}_{12} > 0$), indicating antiferromagnetic interaction between the nearest magnetic atoms (See Fig. S6, Sec. II) \cite{48}.  

\begin{figure*}[th!]     
\centering           
\includegraphics[width=16cm,height=8cm]{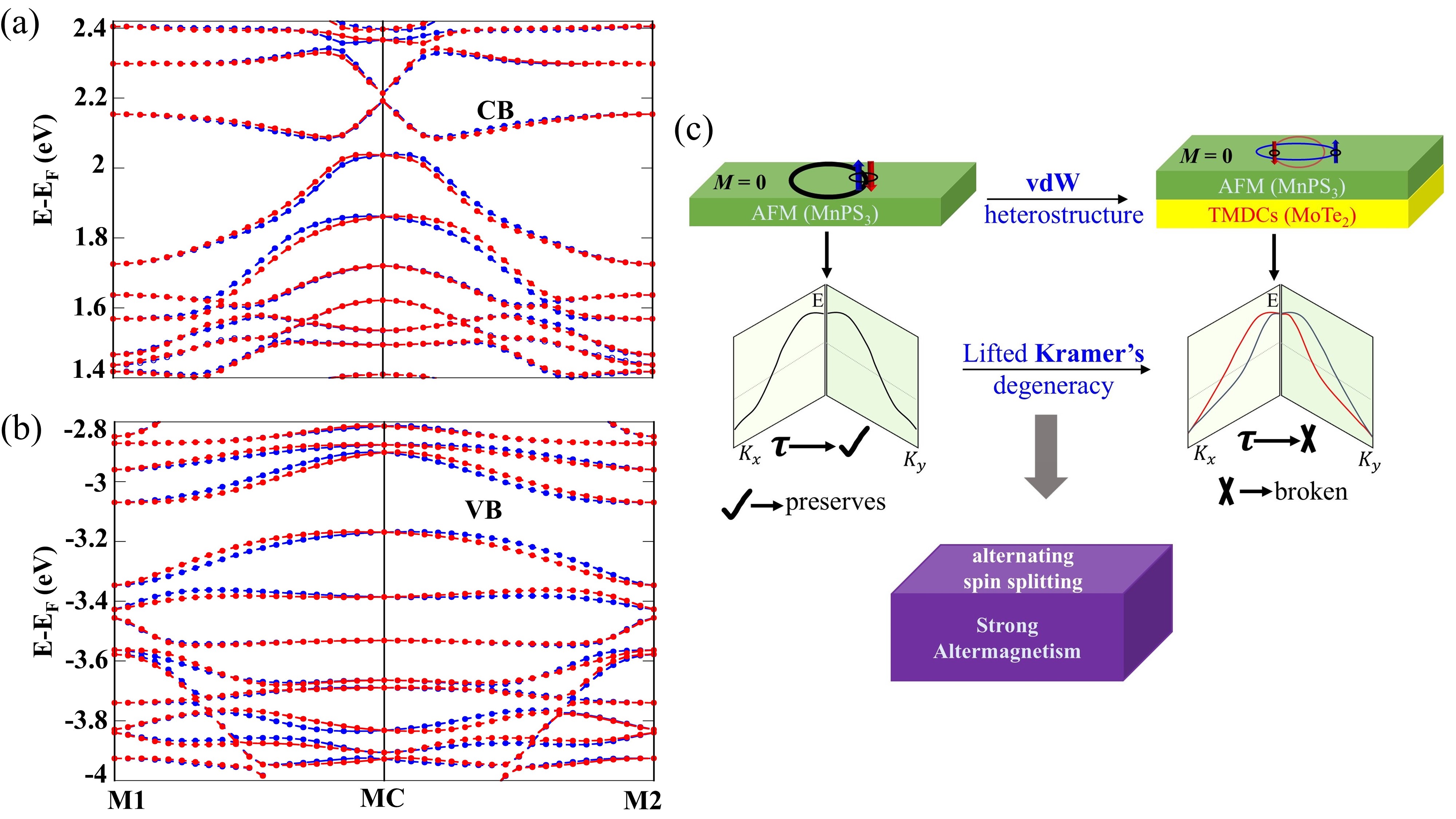}
\caption{\label{Fig:wide}  Electronic structure and altermagnetism without spin-orbit coupling. Bandstructure in (a) conduction band and (b) valence band region, (c) schematic representation of the emergent altermagnetic phase in the designed Mn${}_{2}$P${}_{2}$S${}_{6}$/Mo${}_{3}$Te${}_{6}$ vdW heterostructure. The blue and red color in bandstructure depicts spin-up and spin-down states respectively. }
\end{figure*}
The AM observed in the vdW heterostructure is presented schematically in Figure 3(c). It describes the broken TRS, which leads to a non-degenerate energy spectrum and signifies transition from an in-plane antiferromagnet to an altermagnet. Further, the calculated density of states (DOS), as depicted in Fig. 4(a), confirms the AFM nature of the material, where the spin-up states are compensated by the spin-down states, resulting in zero net magnetization. The DOS curve manifests the semiconducting nature with a band gap of 1.48 eV. Noticeably, the states around VBM and CBM are contributed by MoTe${}_{2}$ layer, whereas the states away from Fermi energy can be attributed to the ML-MnPS${}_{3}$. The projected density of states (PDOS) as shown in Fig. 4(b), predicts spin up states of Mn1 atoms compensated via spin down states of Mn2 atoms and vice versa. It confirms the symmetry-compensated alternating spin-splitting observed in the bandstructure. The hybridization between atomic orbitals demonstrate SOC effect alongwith non-relativistic spin splitting (NRSS). The observed orbital hybridization and size of the alternating spin splitting depend on interlayer interactions. Since the vdW interaction disappears with increasing interlayer distance, the spin splitting correspondingly decreases. To visualize the effect of interlayer interactions and tune the size of spin splitting, we designed and optimized Mn${}_{2}$P${}_{2}$S${}_{6}$/Mo${}_{3}$Te${}_{6}$ vdW heterostructure at interlayer separations of 4.8 Å and 5.7 Å  [refer to Fig. S2]. The effect of interlayer interactions on the bandstructure is shown in Fig. 4(c-f). Intriguingly, the alternating spin splitting at large interlayer separations of 4.8 {\AA} [Fig. 4(c, d)] and 5.7 {\AA} [Fig. 4(e, f)] decreases significantly compared to spin splitting at 3.3 {\AA} [refer to Fig. 3(a, b)].

 \begin{figure*}[th!]                
\includegraphics[width=16cm,height=8cm]{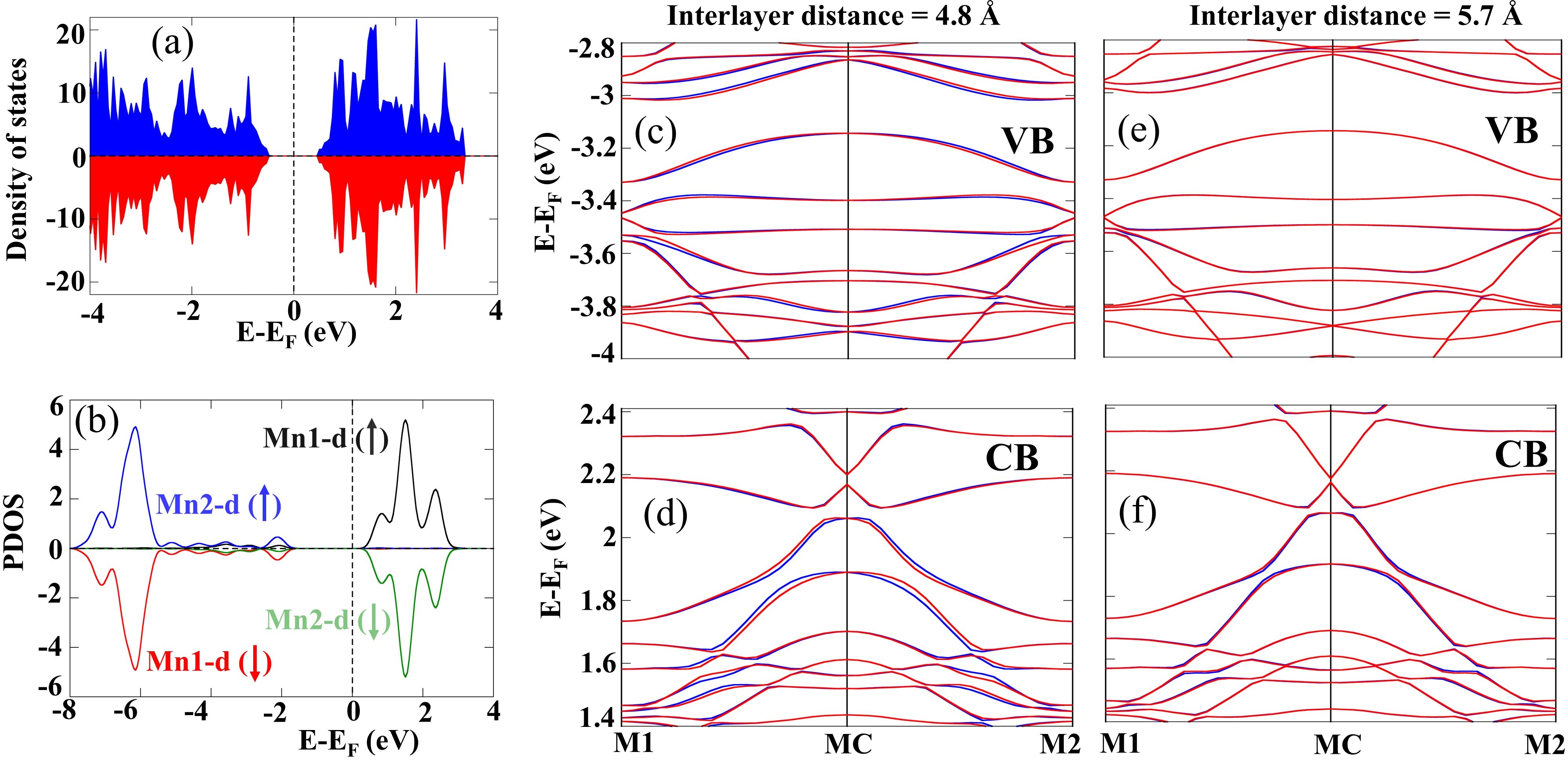}
\caption{\label{Fig:wide}  (a) Spin-polarized density of states, (b) Projected density of states (PDOS), (c, d) bandstructures at interlayer separation of  4.8 {\AA} and (e, f) 5.7 {\AA} respectively [Effect of interlayer interactions on spin splitting]. The red and blue color in Fig. (a, c-f) represents spin-up and spin-down states, respectively. }
\end{figure*}

The presence of SOC leads to the gap opening while preserving the alternating NRSS, which can be attributed to the orbital hybridization. The non-relativistic spin degeneracy is lifted even in the presence of SOC, demonstrating the appearance of strong AM. The wannierized electronic states, as depicted in Fig. 5(a), reveal an identical bandstructure to that in Fig. 3(a, b). The alternating NRSS is intact upon incorporation of SOC, which is apparent in Fig. 5(b) [zoom-in of the shaded region of Fig. 5(a)]. It reveals evidence of altermagnetic states along MC$\mathrm{\to}$M1/M2 path, with lifted Kramer's degeneracy in the reciprocal space, stemming from the locking of spins to their linear momentum. The color bar [as shown in Fig. 5(a, b)] clearly shows that the spin-up ($+1$) states along M1 are compensated by the spin-down ($-1$) states along M2, resulting in zero net magnetization. Additionally, it reveals the energy-dependent spin states along the high-symmetry path, which can be inferred from the varying magnitudes of spins ranging from $-1$ to 1. Intriguingly, the calculated bandstructure in the presence of SOC upholds the intrinsic electronic states while opening a sizeable gap and avoiding band crossings, as shown in Fig. 5(c). This is clearly shown in Fig. 5(d) [zoom-in of the shaded region of Fig. 5(c)]. This large band gap opening can be attributed to the lifting of the band degeneracy due to the interaction between spins and orbital motion of electrons. Notably, the observed bandstructure, both with and without SOC, suggests large alternating spin splitting with zero magnetization. This demonstrates the robustness of the altermagnetic phase in the designed vdW heterostructure. Therefore, the observed altermagnetic behavior can persist from the bulk to the surface states, potentially contributing to the topological properties as well.

\begin{figure*}[th!]                
\includegraphics[width=16cm,height=8cm]{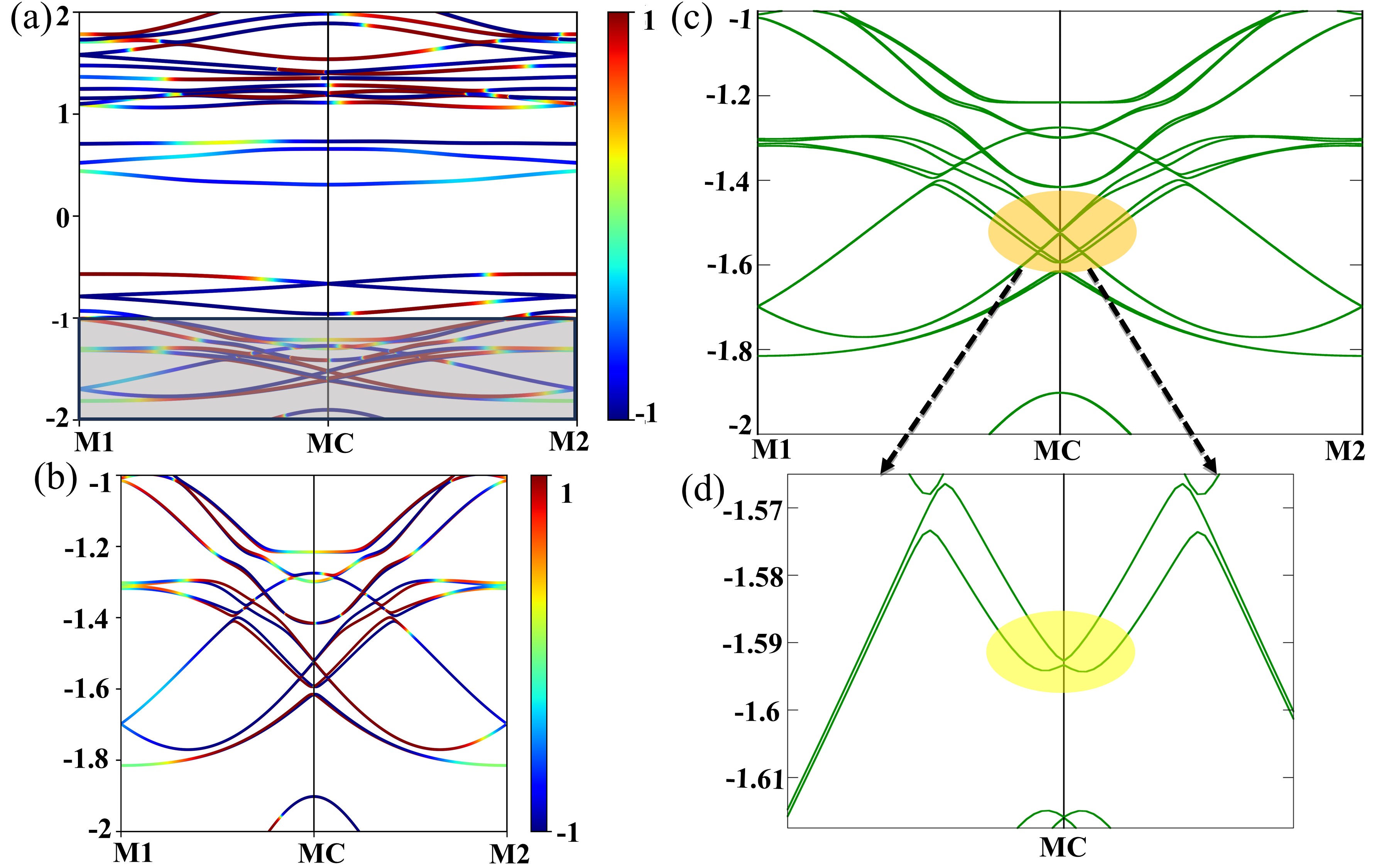}
\caption{\label{Fig:wide}   Effect of spin-orbit coupling on bandstructure. (a, b) Spin-polarized wannierized bandstructure [(b) represents bandstructure in valence band region] and (c, d) unpolarized bandstructure in presence of spin-orbit coupling [(d) represents gap opening in the shaded region]. The color bar in Fig. (a, b) represents the expectation value of the spin \textit{Z}-component. }
\end{figure*}

Further, the observed TRS breaking and significant spin polarization can be correlated with the non-zero Berry curvature. The Berry curvature in the first Brillouin zone along the high-symmetry points, as shown in Fig. 6(a), is determined following the Kubo formula as $\mathrm{\Omega}_n^Z\left(K\right) = - \sum_{m \neq n} 2 \, \text{Im} \left(\langle \psi_{nk} |v_x| \psi_{mk} \rangle \langle \psi_{mk} |v_y| \psi_{nk} \rangle \right) \frac{1}{\left( \omega_m - \omega_n \right)^2}$ where, $\omega_{m/n} = \frac{E_{m/n}}{\hbar}$ and $v_{x/y}$ represents velocity operators \cite{49}. Consequently, the total Berry curvature of Bloch electrons in occupied states is obtained using $\mathrm{\Omega}^Z(K) = \sum_{n} f_n \mathrm{\Omega}_n^Z\left(K\right)$, where $f_n$ corresponds to the Fermi-Dirac distribution \cite{50}. The finite magnitude of the Berry curvature, along with opposite signs, demonstrates the breaking of TRS in the heterostructure through local spin splitting (polarization). Such a non-zero Berry curvature with zero magnetization can significantly influence the transport of charge carriers with different spin states. Consequently, the Berry curvature exhibiting such exotic behavior can govern the unconventional AHE in the designed altermagnetic system. The aforementioned observations suggest an AHE as a consequence of lower crystal-symmetry-derived alternating spin splitting and TRS breaking. The AHC as shown in Fig. 6(b) unravels the extraordinary magnetotransport properties of the designed altermagnetic material. It is estimated by integrating the Berry curvature throughout the Brillouin zone using $\sigma_{xy} = -e^2 \hbar \int_{\text{BZ}} \frac{d^3k}{(2\pi)^3} \Omega_Z(k)$, where e is the electronic charge and $\hbar$ represents the Planck constant \cite{50}. The maximum value of AHC is found to be $\mathrm{\sim}$732.9 S/cm at -0.94 eV. This large AHC stems from non-zero Berry curvature [refer to Fig. 6(a)], in addition to the non-relativistic effects. The inset of Fig. 6(b) represents drifting of electrons with opposite spins away from each other, leading to the observed AHE in the modeled altermagnet. To achieve this energy level (~ -1 eV) below the valance band maximum, hole doping can shift the Fermi level and makes it possible to harness the observed AHC behavior for practical purposes. In this regard, hole doping in the designed system leads to the shifting of Fermi level well below valance band maximum, as shown in Fig. S3. Figure S3(a-c) shows the band structure at doping concentrations $3 \times 10^{14} \,\text{cm}^{-2}$, $4 \times 10^{14} \,\text{cm}^{-2}$ and $5 \times 10^{14} \,\text{cm}^{-2}$, respectively. It is clearly seen that Fermi level shifts (denoted by arrows) with increasing hole doping concentration and at $5 \times 10^{14} \,\text{cm}^{-2}$, it shifts to the desired energy value (at which maximum AHC is observed). It is worthwhile to mention that the estimated carrier concentration to shift the Fermi level towards valance band region is practically achievable \cite{51,52,53,54}. Additionally, the outstanding magnetotransport behavior can be attributed to the inequivalent Mn atoms with opposite magnetic moments, which are connected by rotational symmetry. The reduction of crystallographic symmetry in the vdW heterostructure allows for the formation of an axial Hall vector, which is crucial for a non-vanishing AHC and is odd under time reversal \cite{55,56}.

\begin{figure*}[th!]     
\centering           
\includegraphics[width=16cm,height=8cm]{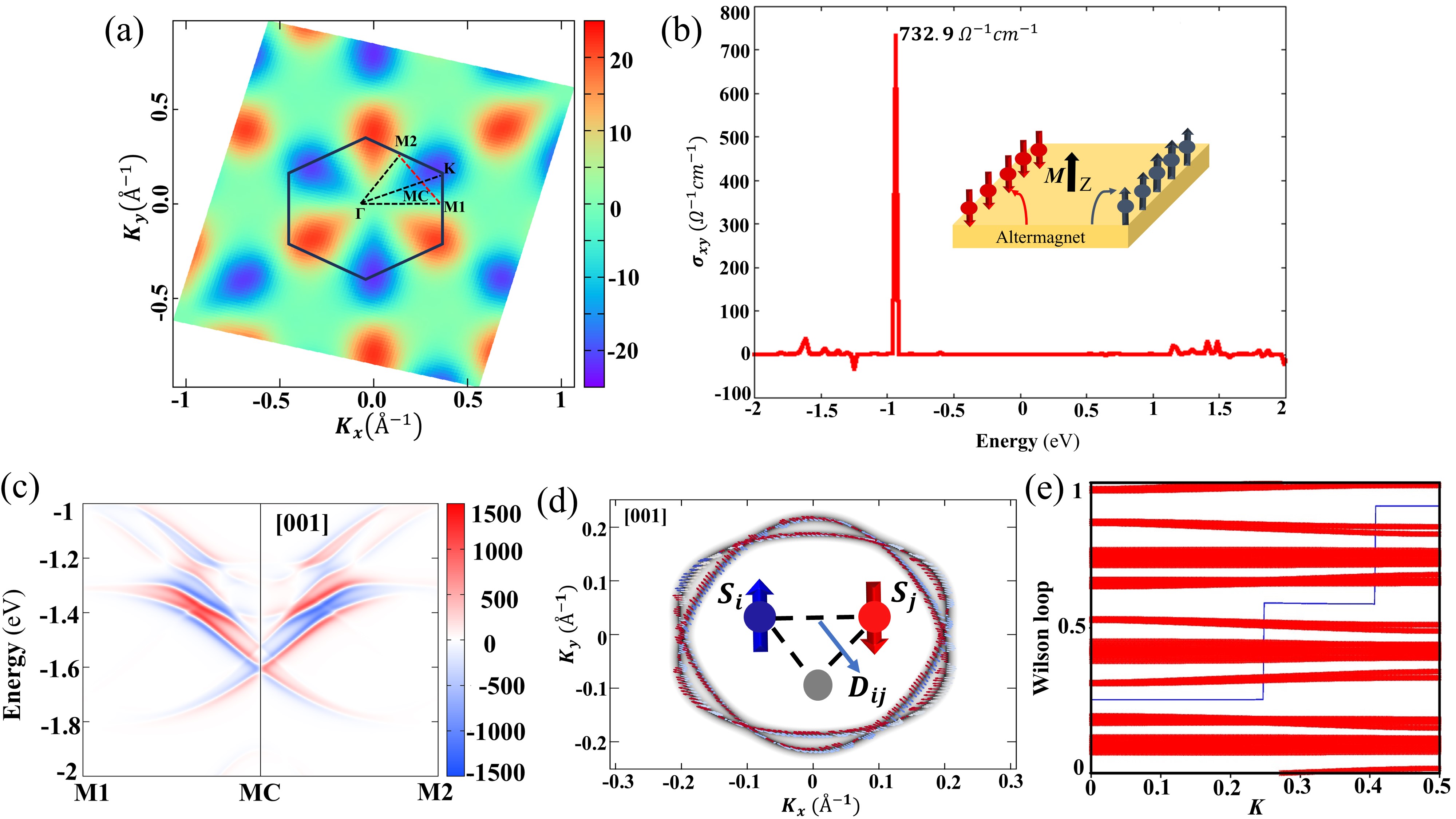}
\caption{\label{Fig:wide}  Berry curvature, magneto-transport (unconventional AHE) and its topological behavior. (a) Non-zero Berry curvature in the momentum space, (b) anomalous Hall conductivity [Inset: Schematic representation of anomalous Hall effect], (c) surface states with alternating spin splitting along [001], (d) spatially inhomogeneous spin texture [Inset: Dzyaloshinskii-moriya interaction] and (e) Wannier charge center [blue line crosses the Wannier states odd time] along [001] principal plane respectively. The color bar in Fig. (a) represents the magnitude of the Berry curvature in the first Brillouin zone, whereas in Fig. (c), the red and blue color represents spin-up and spin-down states, respectively. }
\end{figure*}

Moreover, spin-polarized surface states along definite principal planes can give rise to nontrivial topological characteristics. This crystal symmetry allowed planes, host surface states with alternating spin splitting along the specified high-symmetry path. In line with this, the surface states corresponding to the \textit{S${}_{z}$} component of spin in the VB region exhibit alternating spin splitting in the momentum space, as shown in Fig. 6(c). This aligns with the predicted altermagnetic characteristics and confirms the existence of the observed bulk compensated spin splitting bands in the surface states as well. Notably, the observed altermagnetic phase persists along the [001] plane, whereas the altermagnetic spin splitting annihilates along the [010] and [100] principal planes [Fig. S4] \cite{48}. This can be inferred to crystal symmetry-enforced spin degeneracy, resulting from merging of high-symmetry points of the opposite sublattices, leading to the compensated collinear magnetic configuration along [010] and [100]. However, in contrast to the alternating surface states [refer to Fig. 6(c)], the bulk energy band spectrum along [001] principal plane as shown in Fig. S5 (left panel) is found to be trivial in nature \cite{48}. It is analogous to the bandstructure presented in Fig. 5(b). However, the celebrated transport properties can be correlated with the momentum-dependent spin arrangements. Figure 6(d) reveals that the spins are entangled to the motion of electrons in the momentum space. Intriguingly, the spin texture presents spatially inhomogeneous distribution of spins, resulting in vanishing total magnetization. The vanishing net magnetization indicating antiferromagnetic ground state, can be attributed to a smaller out-of-plane DMI ($\bm{D}_{12}^{\bm{z}}$) value of 3.7 meV, originating from the SOC of Rashba-Dresselhaus type and hybridization  between p and d orbitals. The designed system exhibits altermagnetic behavior due to crystal symmetry and time reversal symmetry breaking via vertical vdW interactions, which is along Z direction. Consequently, in a vertically stacked vdW heterostructure, wherein the ground state altermagnetic configuration stems from symmetry breaking, the vertical components of exchange interaction play crucial role. Herein, the vertical component of DMI exchange interaction could dominate over in-plane components, since the designed heterostructure is vertically stacked and also the anti-parallel atomic magnetizations are oriented along Z-direction. Hence, the vertical DMI component is the key parameter that determines the strength of SOC and the size of spin canting. Here, the out-of-plane DMI is estimated using four states method as mentioned in supporting information under Sec. II [refer to Fig. S7] \cite{48}. Noteworthy, the ratio of $\frac{\bm{D}_{\mathbf{12}}^z}{\bm{J}_{\mathbf{12}}}=0.23$, which indicates smaller DMI with respect to Heisenberg interation. Hence, the observed weak out-of-plane DMI interaction suggests small spin canting and stabilizes collinear antiferromagnetic ground state in the designed inversion symmetry broken vdW altermagnet. Specifically, the spin texture exhibit Dresselhaus SOC with minimal Rashba effect. Additionally, the spins are entangled to its momentum at right angles (i.e. 90$\mathrm{{}^\circ}$), which confirms the emergence of spin-momentum locking in the system. Such spin-momentum locking favors nontrivial topological features, besides magnetotransport properties in the altermagnetic phase. The Fermi arc, demonstrates nodal points where the spin degeneracy is preserved, as the wavefunctions of spin-up and spin-down electronic states overlap in momentum space [refer to Fig. S5] \cite{48}. The hallmarks of SOC-endowed surface states, including alternating spin splitting, spin texture, and non-relativistic spin splitting indicate nontrivial topology. In this regard, we calculated Wannier charge center (WCC) as shown in Fig. 6(e), which corresponds to the Wilson loop and manifests nontrivial band topology. The crossing of step-like blue line across the Wannier states (an odd number of times)  confirms the emergence of nontrivial topological behavior in the system. Further, the non-zero Chern number, which is found to be 1 (odd) corroborates the observed topological properties. This makes the modeled altermagnetic vdW heterostructure an efficient candidate for realizing unconventional AHE and nontrivial topology with multifunctionalities.

\section{\textbf{CONCLUSION}}
In summary, strong altermagnetism with unprecedented alternating spin splitting and magneto-transport has been demonstrated in a vdW heterostructure. The broken space-spin symmetry enforces lifting of Kramer's degeneracy, which leads to the unconventional breaking of the time reversal symmetry. The designed Mn${}_{2}$P${}_{2}$S${}_{6}$/Mo${}_{3}$Te${}_{6}$ heterostructure with excellent thermal stability exhibits the spin-orbit interaction alongwith non-relativistic effects, which preserves the alternating bandstructure. This leads to the non-zero Berry curvature and favors outstanding AHE, which is estimated to be $\sim 732.9\ {\mathit{\Omega}}^{-1}{cm}^{-1}$ at -0.94 eV. It can be achieved through shifting the Fermi level below the valence band maximum via hole doping. Notably, the space-spin symmetry of the heterojunction endorses surface-dependent altermagnetic behavior. In this regard, the principal [001] plane preserves the altermagnetic spin splitting, whereas altermagnetism is annihilated on the [100] and [010] planes due to the merging of high-symmetry K-points with opposite spins. Additionally, the observed spin-momentum locking leads to nontrivial topological behavior, as indicated by the non-zero Chern number. Therefore, the manifestation of non-relativistic altermagnetism in vdW heterostructures with exceptional magnetotransport suggests a unique methodology, which expands the landscape of magnetism-driven low-dimensional quantum phenomena.

\section*{\textbf{ACKNOWLEDGMENTs}}
The authors would like to acknowledge Centre for Development of Advanced computing, Pune and Indian Institute of Technology Guwahati for providing computational facilities under the National supercomputing mission. We appreciate the valuable inputs from Saransha Mohanty, Liyenda Gogoi and Deepshekhar Roy.

\bibliography{manuscript}% Produces the bibliography via BibTeX.

\end{document}